\documentclass[preprint,showpacs,preprintnumbers,nofootinbib]{revtex4}

\usepackage{mathtext}
\usepackage[english]{babel}
\usepackage{epsfig,amsmath,amssymb,amsfonts}
\usepackage{bm}
\usepackage[usenames]{color}

\usepackage{graphicx}
\usepackage{mathrsfs}
 \usepackage{psfrag}
\usepackage{rotating}
\usepackage{multirow}

\newcommand\de{\delta}

\newcommand\om{\omega}

\newcommand\De{\Delta}

\newcommand\Lam{\Lambda}

\newcommand\half{{\frac{1}{2}}}

\newcommand\lam{\lambda}

\newcommand\RE{{\rm Re}}
\newcommand\IM{{\rm Im}}

\newcommand\quat{{\frac14}}

\begin{document}
\title{TO THE NATURE OF NUCLEAR FORCE}
\author{V.I. Kukulin}
\email{kukulin@nucl-th.sinp.msu.ru}
\author{V.N. Pomerantsev}
\email{pomeran@nucl-th.sinp.msu.ru}
\author{O.A. Rubtsova }%
\email{rubtsova-olga@yandex.ru}
\author{M.N. Platonova }%
\email{platonova@nucl-th.sinp.msu.ru}

\affiliation{%
Skobeltsyn Institute of Nuclear Physics, Lomonosov Moscow State
University, Leninskie Gory 1/2, 119991 Moscow, Russia}


\begin{abstract}
It has been shown for the first time that $NN$ interaction, at least in some
partial waves, can be quantitatively described by the superposition of a
long-range one-pion exchange and a short-range mechanism based on the complex
pole in the $NN$ potential corresponding to the dibaryon resonance in this
partial wave. For the partial waves $^3P_2$, $^1D_2$, $^3F_3$ and $^1S_0$ the parameters of the complex poles that give the best description of the elastic and inelastic phase shifts of $NN$ scattering are very close to the empirical parameters of the corresponding isovector dibaryon resonances detected experimentally. Based on the results obtained, a general  conclusion is made about the nature of nuclear force at medium and small internucleon  distances.
\end{abstract}


\maketitle

\bigskip
\begin{center}
1. SHORT INTRODUCTION TO THE PROBLEM
\end{center}

 The problem of nuclear force has been the subject of so many works in the scientific literature since the 1930s, that there is no possibility within a single paper to even list all the models and approaches proposed in this area. Therefore we refuse such an attempt, leaving this task for an appropriate review or monograph (a far from complete consideration of existing nuclear force models is contained in the book~\cite{Book} and the review~\cite{Machleidt}), and will briefly discuss only works closest in topic to the approach proposed here.

Almost all theoretical approaches proposed so far to describe nuclear forces are anyway based on the classical Yukawa concept, in which the main carrier of the strong internucleon interactions in nuclei is the meson exchange between nucleons (or between quarks constituting nucleons) in the $t$ channel, generating the transferred momentum singularities of the scattering amplitude. However, it should be noted that this basic concept meets so many internal contradictions and difficulties when trying to describe experimental data, that the question arises inevitably about the correctness and application limits of the meson-exchange picture of nuclear forces.

Let's quote here the opinion of the well-known American physicists working in this field (see \cite{Barnes}) about the vector meson exchange between nucleons which, according to the traditional point of view, is responsible for the strong short-range repulsion (the so-called repulsive core):
``A literal
attribution of the short-range repulsive core to vector meson
exchange, as opposed to a phenomenological parametrization, of
course involves a {\em non sequitur}: since the nucleons have
radii $\approx$ 0.8 fm and the range of the vector exchange force
is $\hbar/m_\omega c\approx 0.2$~fm one would have to superimpose
the nucleon wavefunctions to reach the appropriate internucleon
separations. The picture of distinct nucleons exchanging a
physical $\omega$-meson at such a small separation is clearly a
fiction...''.
Similar problems arise also when considering other types of meson exchanges: scalar, pseudoscalar, etc. (for more details on the existing problems and contradictions in traditional models of nuclear forces, see \cite{Holinde,Gloeckle96,YAF2013}).

Now the dominant approach to the quantitative description of nucleon-nucleon interaction is the so-called Effective Field Theory (EFT) or Chiral Perturbation Theory (ChPT) with respect to the small parameter $Q/\Lam_{\rm QCD}$, where
$\Lam_{\rm QCD} \sim 1$~GeV is a characteristic momentum parameter of Quantum Chromodynamics, and $Q$ is the momentum transferred in the course of interaction~\cite{Weinberg, Machl, Eppelbaum, Ordones}.
In this approach, the peripheral part of $NN$ interaction is described via the superposition of the  perturbative series terms in the successive orders: leading order (LO), next-to-leading order (NLO), next-to-next-to-leading order (N$^2$LO), etc., whereas all short-range contributions are parameterized via the so-called contact terms, which, according to this concept, {\em should not depend} on energy, as well as on the approximation order.
By construction, this general approach should be valid only up to the collision energies $T_{\rm lab} \simeq 350$~MeV, where the expansion parameter
$Q/\Lam_{\rm QCD}$ remains still relatively low, i.e., up to energies a little bit over the  pion-production threshold. At higher collision energies, the EFT approach should be supplemented by some model able to describe properly the short-range components of $NN$ interaction and the corresponding short-range correlations in nuclei.

As such a supplement, one can consider the well-known quark model and its various versions known today.
It should be said that some attempts to treat short-range
$NN$ force within the quark model
were undertaken since the late 1960s (see,
e.g., works of Yazaki et al.~\cite{Yazaki}, Miller et al.~\cite{Miller}, Faessler et al.~\cite{Faess}, Neudatchin et al.~\cite{Neudat} and the works of many other groups).
Unfortunately, such a hybrid treatment of $NN$ interaction inevitably leads to the problem of double counting, because in quark models the one-gluon exchange is usually supplemented by meson ($\pi$ and $\sigma$) exchanges between quarks, which immediately leads to the appearance of the corresponding meson-exchange forces between nucleons not only at short, but also at medium and long distances. In addition, in such hybrid approaches there arise complicated problems with the exchange of scalar and vector mesons between quarks (see the above citation from \cite{Barnes}).
The main problems with the consistent quark-model approach are clearly seen in the work~\cite{Stancu}, where the authors used the quark-quark interaction best fitted to the spectra of excited nucleons. It was shown that if one takes $qq$ interaction in the form of the so-called Goldstone-Boson Exchange (GBE), then one gets a purely repulsive interaction in the $NN$ sector \cite{Stancu}. These difficulties gave rise to general skepticism regarding the possibility of a quantitative description of $NN$ interaction within the framework of microscopic quark models.

In the light of the foregoing, it would be highly desirable to describe $NN$ interaction at short and medium distances without involving full microscopy of the quark model (which is still poorly known today), but using some QCD-motivated models that reproduce correctly the main effects of the six-quark system in various partial waves, but without a detailed description of the whole complexity of multiquark dynamics.

In the present authors' opinion, such objects, which are closely related to the the six-quark dynamics, on the one hand, and can decay (virtually or really) into
$NN$, $N\Delta$ and $\Delta\Delta$ channels at relatively low energies, on the other hand, can be dibaryon resonances, which were predicted by Dyson and Xuong as early as 1964~\cite{Dyson}, at the dawn of the quark era.
It should be specially emphasized that the first dibaryon resonance was experimentally discovered as early as the mid 1950s, long before the first work on the quark model of hadrons, in the works of the Meshcheryakov group in Dubna~\cite{Meshcher}, carried out before the organization of JINR in 1956. It is also important to add that the mass of the isovector dibaryon found in these experiments was employed as an energy scale of the $SU(6)$-multiplet splitting for theoretical prediction of dibaryon masses by Dyson and Xuong~\cite{Dyson}.
Just recently, after many years of rejection, doubts, and unreliable findings, some of the previously predicted dibaryon resonances have finally been reliably discovered in experiments~\cite{Bashkanov09, Adlarson11, Adlarson14, Adlarson18, Komarov} (see also the recent review~\cite{Clement}),
and their decays into $NN$, $ N\De$ and $\De\De$ channels have been studied.

Dibaryon resonances are very attractive for explaining the properties of short-range $NN$, $N\De$ and $\De\De$ forces not only due to their six-quark structure but, first of all, because they are relatively long-lived states formed in the $NN$ system, and six-quark dynamics in such states is manifested most clearly.

We should say here a few more words about the quark-model approach to the problem of $NN$ interaction in general. In the above cited works
~\cite{Yazaki, Miller, Faess, Neudat, Stancu}, the resonating group method (RGM) (known in nuclear physics) was chosen as the main approach to the description of six-quark dynamics. This method assumes that the three-quark wave functions taken for isolated nucleons are not distorted by $NN$ interaction, and there is only a direct exchange of quarks between interacting nucleons. As input $qq$ interactions, the potentials of one-gluon and one-pion exchange were used, which were later supplemented by the purely phenomenological potential of the light scalar meson ($\sigma$) exchange between quarks.
Success of such microscopic approaches, even for the description of only elastic $NN$ scattering, occurred to be rather modest, and therefore they gave way to more well-founded dynamic hybrid approaches.

First of all, it is worth mentioning the Quark Compound Bag
(QCB) model developed by Simonov et al. in the early 1980s~\cite{Simonov, Kal, GKN}, in which the internal six-quark bag is coupled with the external $NN$ channel via contact interaction at the bag boundary.
In this model, the bag size corresponded not to the nucleon r.m.s. radius or the radius of the repulsive core $r_c$, but to the specific radius of the pion exchange $r_{\pi} = \hbar/(m_{\pi}c)$, and the basic six-quark configuration of the bag was assumed to be symmetric one corresponding to an unexcited state.
The energy scale in such a bag ($\sim 600$~MeV) was determined by the position of the poles of the $P$ matrix, which is the inverse of the $R$ matrix in the Wigner theory of reactions. The coupling of the quark bag to the $NN$ channel was defined by matching the internal and external wave functions at the bag boundary similarly to the $R$-matrix theory.

Further mathematical formalism for such a hybrid model was developed in the works of the Leningrad group~\cite{Merkur}, in which a two-channel approach with internal and external channels (without matching of wave functions at the bag boundary) has been proposed. In this approach, dynamics of a six-quark bag in the internal channel was included in the usual quantum-mechanical Hamiltonian scheme by extending the Hilbert space. (Note that the approach proposed by the present authors employs the same elegant mathematical scheme of the Leningrad group.)

Another version of the same idea was developed by Feshbach and Lomon in their boundary conditions model\cite{Feshb-Lomon}. In this model, the internal wave functions of the six-quark bag are matched at the boundary with an external $NN$ channel, where the interaction is governed by a one-meson exchange similar to the traditional one-boson exchange potentials.

However, since the publication of all these works, a very large number of new experimental and theoretical results have been accumulated in the field of nuclear and hadron physics that allow us to look at the whole problem of $NN$ force from a completely different position.

First, as we noted above, a few dibaryon resonances have been reliably discovered, and the basic  properties of their decay into different channels have been studied~\cite{Bashkanov09, Adlarson11, Adlarson14, Adlarson18, Komarov, Clement}.
Second, it turned out that the known dibaryon resonances are strongly coupled to inelastic channels, while their coupling to the elastic $NN$ channel is much weaker (this can be illustrated by comparing the decay widths of the dibaryons into the elastic and inelastic channels: $\Gamma_{\rm inel} \gg
\Gamma_{\rm el}$).
Third, in numerous experiments on scattering of high-energy electrons off different nuclei, performed in recent years at the Jefferson National Laboratory~\cite{J-Lab1, J-Lab2, J-Lab3, J-Lab4}, it was shown that in nuclei there exist tightly correlated pairs and triples of nucleons at short $NN$ distances, that scatter fast electrons like densely packed quark clusters. The momentum distributions of quarks within such clusters are directly related to the nature of short-range nucleon correlations in them~\cite{Hen}.
And finally, starting from the 1970s, the so-called cumulative processes caused by interaction of incident fast hadrons with dense nucleon clusters in nuclei have been studied in detail, and the phenomena of nuclear scaling and superscaling have been discovered~\cite{Baldin, Leksin}.

It is evident that the picture of short-range correlations of nucleons in nuclei emerged from these experimental results is very far from traditional ideas about the universal repulsive $NN$ core in the spirit of Jastrow (for more details, see~\cite{YAF2013}).
So it becomes quite obvious that the interaction of nucleons at short and intermediate distances is inextricably linked to their quark structure and, especially, to the relatively long-lived di- and multibaryon resonances. Today one can ask the following question: what is the real impact of the known dibaryon resonance in a given $NN$
partial channel ${}^{2S + 1}L_J$ on the behavior of elastic and inelastic phase shifts of $NN$ scattering in this channel? In this regard, it should be emphasized that, in contrast to previous works in the field, {\em we consider both elastic and inelastic channels of nucleon interaction
simultaneously} (i.e., within the framework of the same model).

Thus, the aim of this work is to prove that many, if not all, partial channels of $NN$ interaction can be described by a superposition of the long-range potential of one-pion exchange and one complex pole corresponding to the experimentally observed dibaryon in the channel under consideration. Keeping in mind that the impact of one-pion exchange and its role in the full $NN$ interaction are relatively moderate, one can conclude that the dominant role in $NN$ interaction (in the given partial waves) is played by the $s$-channel dibaryon resonance exchange.

The structure of the work is as follows. In Sec. 2 we present a qualitative picture of $NN$ interaction, which follows from the dibaryon model of nuclear forces proposed by one of the present authors in 1998 \cite{PIYAF}. Here we explain how the repulsive core at short distances appears in such a model and what is the origin of the basic attractive force at intermediate distances.
In Sec. 3 the formalism used in the work to describe both elastic and inelastic phase shifts of $NN$ scattering is given. Here we derive the effective separable potential corresponding to a single pole in the internal channel. In Sec. 4, we consider the single-pole description of elastic and inelastic $NN$ scattering in isovector ($T = 1$) channels $^3P_2$, $^1D_2$, and $^3F_3$ (isoscalar ($T = 0$) channels will be considered in a separate work). Sec. 5 is devoted to description of the channel $^1S_0$, where there appears not one but two poles in $S$ matrix. In Sec. 6, we discuss the important role that dibaryon resonances play in nuclear physics in general. In conclusion we summarize briefly the novel concept of nuclear force described in the paper.

\bigskip
\begin{center}
2. DIBARION MODEL FOR NUCLEAR FORCE. A QUALITATIVE CONSIDERATION.
\end{center}

In this section, we briefly describe an alternative interpretation within the framework of the dibaryon model of those basic effects of nuclear force that
in traditional approaches of the Yukawa type are described through the exchange of various mesons (scalar, pseudoscalar, vector, etc.) between isolated nucleons.

Note that the previous hybrid approaches combining both quark and meson sectors (such as those of Simonov~\cite{Simonov} or Feshbach--Lomon~\cite{Feshb-Lomon}) were already based on the selected multiquark states, but without any specification of their quark structure and without any correspondence of the employed parameters of quark bags to the experimentally found six-quark resonances (dibaryons).
In addition, all these $R$-matrix-type models inevitably included an arbitrary matching radius of nucleon and quark wave functions, and while in the QCB model \cite{Simonov} this radius was chosen to be quite large ($R_m \simeq 1.4$~fm), in the Feshbach--Lomon model~\cite{Feshb-Lomon} it occurred noticeably smaller
($R_m \simeq 0.9$~fm), that indicated a significant uncertainty in the choice of the matching radius.

One of the most significant drawbacks of these quark models was absence of any connection with the processes of meson production. Therefore, such models were unable to explain the origin of inelastic processes in nucleon collisions. In addition, it was completely unclear what is the physical origin of such effects as traditional repulsive core and strong intermediate-range attraction induced by exchange of the light scalar meson in the traditional force models. As a result, these improved models have also be left by the early 2000s.

The dibaryon model of nuclear forces proposed in the works
\cite{PIYAF, YaF2001, JPhys2001, KuInt, sys3n, AnnPhys2010} succeeded to include quite naturally most of the effects predicted by traditional meson-exchange models, but in a completely new interpretation. The main argument however is that in
the ground of the dibaryon concept there is an experimentally verified assumption about the dominating role of dibaryon resonances in $NN$ interaction in each partial-wave channel.
Besides, the present authors have shown in a series of papers
\cite{Plat2013, NPA2016, PRD2016} that the observed
 dibaryon characteristics are in good agreement with the experimental cross sections for one- and two-pion production in $NN$ collisions at intermediate energies $T_N \sim 1$~GeV.

\bigskip
\begin{center}
\em 2.1. Main features of $NN$ interaction resulted from the
dibaryon force model
\end{center}

1. The first most important point is related to the choice of the short-range cutoff momentum $\Lambda_{mNN}$ in meson-nucleon form factors, which essentially determines the strength of meson-nucleon interaction. The higher the value of the parameter $\Lambda_{mNN}$, the stronger, on average, is the coupling of the given meson with the nucleon and the higher is the momentum that can be transferred from the meson to the nucleon (or vice versa) in the interaction vertex.

Accurate calculations done within the framework of QCD, as well as 
within dynamic models for meson-nucleon scattering, provide quite moderate values of
$\Lambda_{mNN}$ (in particular, $\Lambda_{\pi NN} \simeq 0.5$--$0.9$
GeV/$c$~\cite{Koepf96}), whereas in traditional phenomenological models of $NN$
interaction these values are usually taken strongly increased ($\Lambda_{\pi NN}
\simeq 1.2$--$1.7$ GeV/$c$ --- see, e.g.,~\cite{Machl01}). And only with these
large values for the high-momentum cut-off parameter $\Lambda_{mNN}$, one can
describe empirical $NN$ phase shifts within the framework of traditional
meson-exchange models for $NN$ force.

In sharp contrast to this, in the dibaryon nuclear force model, just the dibaryon exchange (in the $s$ channel) plays the main role in the description of $NN$ scattering, while the meson exchange makes a moderate contribution. As a result, a good description of empirical data is attained already with rather low values of $\Lambda_{mNN}$ (in particular, $\Lambda_{\pi NN}$ and $\Lambda_{\pi N\Delta}$)~\cite{JPhys2001,NPA2016}.

2. The second critical point is related to the origin of the repulsive core in the $NN$ potential. In the traditional approach, the repulsive core is usually associated with the exchange of an isoscalar vector $\omega$-meson with a mass $m_\om\simeq 780$~MeV. In this case, to get the repulsion of the required intensity, one should take the $\omega$-nucleon coupling constant $g^2_{\om NN}/4\pi \simeq 13.6$, while the $SU(3)$ symmetry predicts the value $(g^2_{\om NN}/4\pi)_{SU(3)} \simeq 5.5$, i.e., 2.5 times smaller.
Moreover, Feshbach showed \cite{Feshbach-su3} that the $SU(3)$ symmetry breaking in $NN$ interaction is weak.

\begin{figure}[t]
\centerline{\epsfig{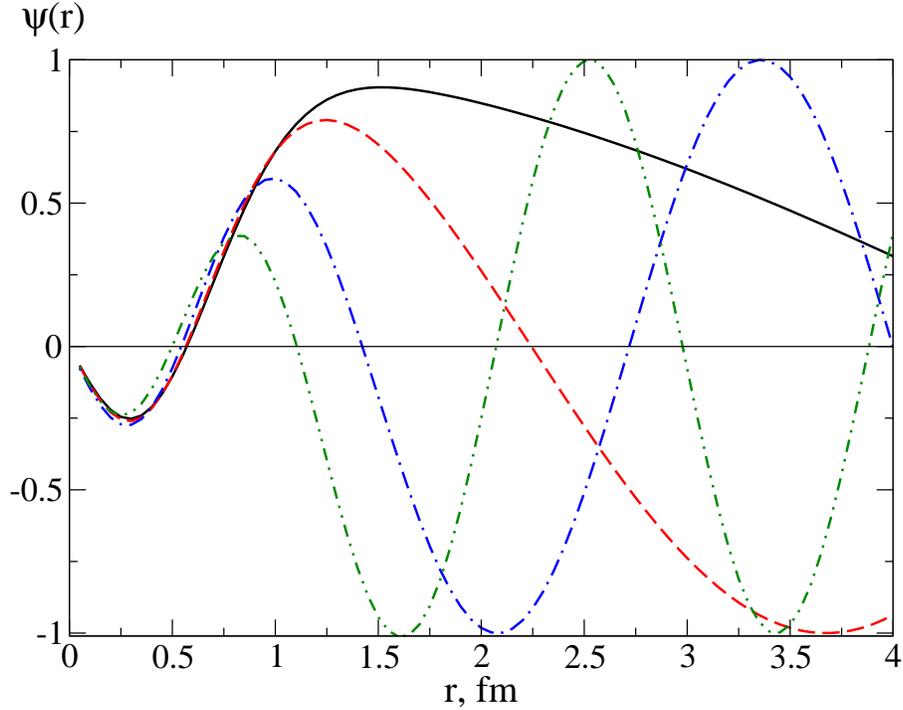}}
\caption{Radial wavefunction of $NN$ scattering in the partial channel $^1S_0$ derived within the framework of the dibaryon force model at various collision energies $T_{\rm lab}$: 10 MeV
(solid curve), 100 MeV (dashed curve), 500 MeV
(dash-dotted curve) and 1 GeV (dash-dot-dotted curve).}
\label{rwf}
\end{figure}

In contrast to these problems, in the dibaryon model
the short-range repulsion arises as a simple consequence of the dominance of the six-quark wave function component with mixed symmetry $|s^4p^2[42]_xLST \rangle$ over the fully space-symmetric component $|s^6[6]LST\rangle$~\cite{YaF2001, JPhys2001, KuInt}.
In fact, the dominant contribution of the mixed-symmetry $6q$ component means the appearance of an internal node in the radial scattering wave functions. This node occurs just at the same position as the traditional repulsive core and is almost independent of energy (see Fig.~1 where the radial wave functions of $NN$ scattering in the $^1S_0$ partial wave at different energies are shown). In Fig.~1 the stationary character of the internal node of the radial wave function is seen very clearly. Thus, it is quite evident that the orthogonality condition to the fully symmetric $6q$ state at short distances plays the same role in the dibaryon model as the strong repulsive core in traditional nuclear force models.

3. The third fundamental question related to the nature of nuclear force concerns with the source of the basic attractive force which holds the nucleons together in a nucleus. In the traditional picture of nuclear force, this attractive interaction is attributed to the exchange by a light scalar $\sigma$-meson with a mass $m_{\sigma}\simeq 450$~MeV and a huge decay width into the $2\pi$ channel $\Gamma_{\sigma}\simeq 550$~MeV \cite{sigma}.
Because of this enormous width, a literal identification of the attractive $NN$ force with a direct $\sigma$-exchange is highly uneasy, since the free path of a particle with such a width $\lam = \hbar c /\Gamma_{\sigma} \simeq 0.4$~fm,
i.e., it is so small (compared to the average distance between nucleons in a nucleus $r_{NN}^{\rm av} \simeq 1.8$~fm) that direct exchange by such a $\sigma$-meson is not able to provide any significant $NN$ attraction. Therefore, in the traditional approach, instead of a direct scalar-meson exchange, one considers a $2\pi$ exchange with two intermediate $\Delta$ isobars \cite{Machleidt} (see Fig.~2).

\begin{figure}[h!]
\centering\epsfig{file=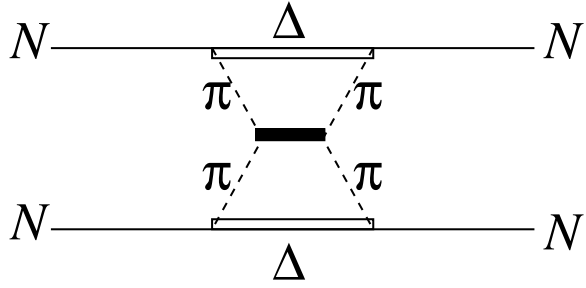,width=0.45\columnwidth}
\caption{$\Delta$-isobar mechanism with two strongly interacting pions in scalar mode in intermediate state, which leads in traditional meson-exchange models to strong attraction at internucleon distances $r_{NN} \simeq 0.7$--$0.9$~fm.}
\label{2pi} 
\end{figure}

However, without strong $\pi\pi$ rescattering in the intermediate state, such a mechanism definitely does not provide an attractive contribution of the required intensity. And since the length of the $\pi\pi$ scattering in the scalar-isoscalar channel is very small ($a_{\pi\pi}\sim 0.2$--$ 0.3 $~fm),
one cannot speak about any strong $\pi\pi$ rescattering of the attractive nature at low energies.
Moreover, it is known that there is repulsion rather than attraction between pions at low energies \cite{Colangelo}. As a result, when using realistic parameters for $\pi N\Delta$ vertices and for $\pi\pi$ interaction at low energies, such a mechanism definitely does not lead to the required attraction.

The dibaryon model provides a completely different explanation of the main attractive nuclear force \cite{YaF2001, JPhys2001, KuInt}. This explanation is based on the three effects listed below.

1) Two nucleons at medium and short distances form a six-quark bag with dominant mixed symmetry $|s^4p^2 [42]_xLST \rangle $. Then it transits (deexcitates) into the completely symmetric configuration $|s^6 [6]_x L'S'T '\rangle
$ (in $S$ waves) with the emission of a scalar $\sigma$ meson, which ``adheres'' to the symmetric bag and strongly constricts it to the center.
In this way, a scalar cloud of virtual $\sigma$ mesons arises around a symmetric quark bag.

2) Due to the fact that the initial state of the $ 6q $ bag
$ | s^4p^2 [42]_xLST \rangle $ is highly excited
(with an excitation energy of
$ 2 \hbar \omega \simeq 500 $--$ 600 $~MeV), which is a simple consequence of the mixed
symmetry of the six-quark wave function in this channel, inside the six-quark
bag, the effect of partial restoration of the QCD chiral symmetry most likely
arises, leading to a significant renormalization and a decrease of the mass of
the $ \sigma $ meson down to near-threshold values $ m_{\sigma}^{\rm ren} \simeq 300 $ MeV (see the works devoted to the chiral
symmetry restoration in excited hadrons~\cite{Glozman, Gassing}, as well as our
work~\cite {Plat2013}). This renormalization of the mass of the scalar meson (together
with a decrease of its width) also effectively enhances the attraction in the $
NN $ channel.

3) As a result of multiple transitions of two nucleons to the state of a dressed six-quark bag and vice versa (see Fig.~3), an effective $NN$ attraction arises induced by the strong coupling with the dibaryon channel. In other words, the system behaves as a two-channel one, where the direct (meson-exchange) interaction in the external (i.e., nucleon-nucleon) channel is rather weak, while the main interaction arises due to strong coupling with the internal (i.e., dibaryon) channel.

\begin{figure}[h!] \centering\epsfig{file=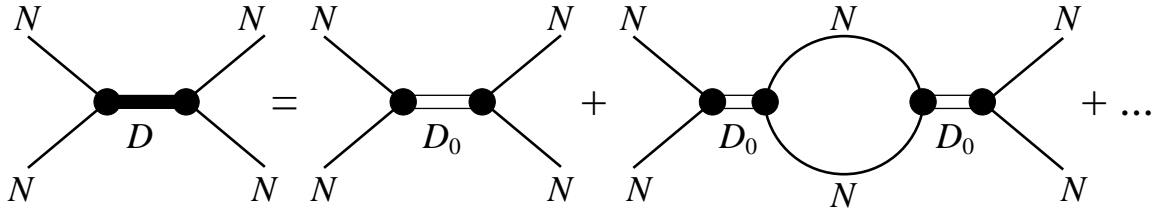,width=0.9\columnwidth}
\caption{``Dressing'' of the total dibaryon propagator $D$ (thick line) with the nucleon loops. Propagator of the ``bare'' dibaryon $D_0$ is shown by a double thin line.}
\label{diagram}
\end{figure}

If we now exclude the dibaryon component and reformulate the problem in the variables of the two-nucleon system only, then a separable $NN$ potential with a coupling constant of the pole type appears (see Sec. 3). This potential at low energies leads to a strong attraction of nucleons, which, in turn, is the main nuclear force.

It is significant in this picture that exactly the same mechanism at collision energies above the two-pion production threshold should lead to inelastic processes of emission of a light scalar $ \sigma $ meson, e.g.,
\[
p+n \to D^* \to d+ \sigma \to d+ (\pi\pi)_0.
\]
Then, in the reactions of two-pion production in the scalar-isoscalar channel, a strongly renormalized intermediate $ \sigma $ meson should appear.

In fact, near-threshold enhancement in two-pion production cross sections in the scalar-isoscalar channel in $NN$, $ Nd $ and $ dd $ collisions was experimentally discovered in the early 1960s and since then it is known as the ABC effect~\cite{ABC}.

Recent high-precision experiments on $ 2\pi $ production not only confirmed the existence of a pronounced ABC effect, but also unambiguously associated it with generation of the intermediate isoscalar dibaryon $ d^*(2380)$~\cite{Adlarson11, Bashkanov2017}.

However, the specific mechanism of the dibaryon decay, leading to the ABC
enhancement, remained unclear until recently. In the work~\cite{Plat2013}, we proposed
the interpretation of the ABC effect as a consequence of the emission of
the renormalized $ \sigma $ meson from the excited dibaryon and showed
that such an interpretation is in good agreement with the experimental
data~\cite{Adlarson11}.

Thus, a strong indication was obtained of the correctness of the dibaryon mechanism of $NN$ interaction through the formation of an intermediate $ 6q $ bag dressed by a field of scalar $ \sigma $ mesons\footnote{%
In contrast, we note that no real evidence for the appearance of  an attractive $ \pi \pi $ correlation has been obtained within the framework of the traditional approach, despite numerous studies of many research groups.}.

\begin{center}
3. DESCRIPTION OF $NN$ SCATTERING INDUCED BY SINGLE STATE IN THE INTERNAL CHANNEL
\end{center}

Below we consider a two-channel model of $NN$ interaction with one complex
pole. It corresponds to the physical picture of $NN$ scattering, governed in
the external channel by the conventional one-pion exchange (OPE), and in the
internal channel --- one six-quark eigenstate with complex energy. The main
difference between the model proposed here and the one presented earlier in
\cite{YaF2001, JPhys2001} is that it is possible to consider both elastic and
inelastic $NN$ scattering due to the presence of an imaginary part of the internal $ 6q $-state energy. We emphasize that the imaginary part of the eigen-energy of this $
6q $ state corresponds to only inelastic (not two-nucleon) modes of its decay,
for example, $ NN\pi $, $ NN\pi\pi $, etc. Thus, we consider a simple
two-channel model in which the first channel corresponds to peripheral OPE
interaction, and the second one --- to a dibaryon resonance with a complex eigenvalue $E_D $.

The full Hamiltonian of such a system has the form:
\begin{equation}
H=\left(
\begin{array}{cc}
h_{NN}& \lam_1|\phi\rangle\langle \alpha|\\
\lam_1 |\alpha\rangle \langle \phi|& E_D|\alpha\rangle \langle
\alpha|\\
\end{array}
\right). \label{Ham}
\end{equation}
Here, the Hamiltonian $h_{NN} = h_{NN}^{(0)} + V_{NN}$ defines the interaction in the external $NN$ channel, which, as we assume, is exhausted by a single-pion exchange, i.e.,
\begin{equation}
V_{NN}=
\frac{f_{\pi}^2}{m_{\pi}^2}\frac{1}{q^2+m_{\pi}^2}\left(\frac{\Lambda_{\pi
NN}^2-m_{\pi}^2}{\Lambda_{\pi NN}^2+q^2}\right)^2({\bm\sigma}_1\cdot
{\bf q})({\bm \sigma}_2\cdot {\bf q})\frac{({\bm\tau}_1\cdot {\bm\tau}_2)}{3}, \label{Vope}
\end{equation}
where $m_{\pi} = (m_{\pi^0} + 2m_{\pi^{\pm}})/3$ and $\Lambda_{\pi NN}$ is the high-momentum cut-off parameter~\footnote {In the calculations, we use the averaged pion-nucleon coupling constant $f_{\pi}^2/(4\pi) = 0.075$ and ``soft'' cutoff $\Lambda_{\pi NN}=0.65$~GeV/$c$.}.

The Hamiltonian of the internal channel in Eq.~(\ref{Ham}) includes the eigen-energy of the dibaryon resonance $ E_D = E_0 -i{\Gamma}_{\rm inel}/2$, the imaginary part of which is determined by
the width of the resonance decay into inelastic channels, $\Gamma_{\rm inel}$.
  In the formula (\ref{Ham}), we introduced also the wave function $|\alpha\rangle$ of a six-quark bag (dibaryon),
the form factor $|\phi\rangle$ for the coupling of the external and internal channels, and the coupling constant $\lam_1$.
The form factor $|\phi\rangle $ is the matrix element of the overlap of the product of the wave functions of two nucleons (obtained in the framework of the quark model) and
the six-quark wave function of the dibaryon in a given partial wave. It is a function of the relative distance (or relative momentum) in the $NN$ channel, and it also depends on the spin, isospin, orbital and total angular momenta of the $NN$ system.
For present calculations we used a simple Gaussian form factor with one parameter $r_0$ \cite{JPhys2001}.

Since the six-quark Hamiltonian (\ref{Ham}) transforms a state vector from the dibaryon channel into the channel describing the relative motion of two nucleons and vice versa, it is convenient to exclude the six-quark space from the complete problem by the usual method~\cite{Feshb}, which gives the effective energy-dependent Hamiltonian in the $NN$ channel:
\begin{equation}
H_{\rm eff}(E)=h_{NN} +\frac{\lam_1^2|\phi\rangle \langle
\phi|}{E-E_D}. \label{Heff}
\end{equation}
Thus, the total interaction in the $NN$ channel consists of the OPE potential (\ref{Vope}) and the separable energy-dependent potential
$\frac{\lam_1^2|\phi\rangle \langle \phi|}{E-E_D}$.
Note that the total scattering amplitude corresponding to the sum of these potentials includes $t$-channel exchange by any number of pions (ladder type) and any number of $NN$ loops, and the nucleons can also exchange by any number of pions inside these loops --- see Fig. 4.

\begin{figure}[h!] \centering\epsfig{file=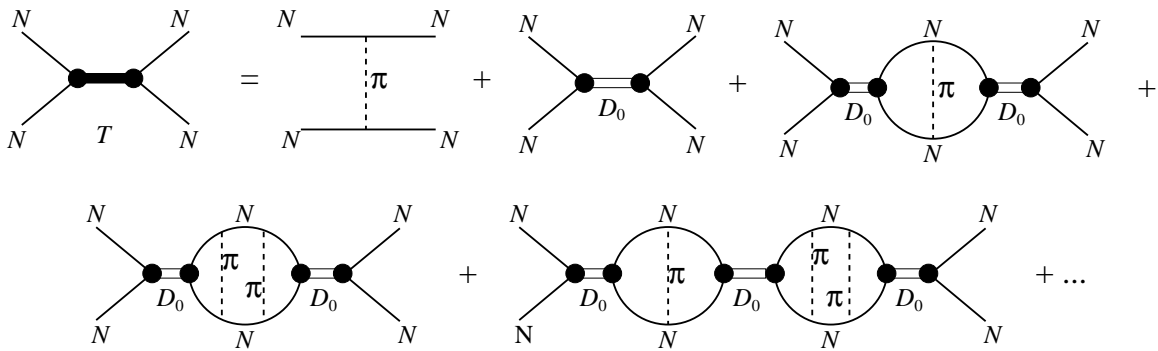,width=0.9\columnwidth}
\caption{Total scattering amplitude for our model, including all possible $t$-channel pion exchanges and transitions to the internal dibaryon channel.}
\label{amplitude}
\end{figure}

Since the main part of the effective Hamiltonian
(\ref {Heff}) has a separable form, one can define an additional  scattering matrix  $ t $ in the distorted-waves representation corresponding to the external Hamiltonian $h_{NN}$. The corresponding transition operator has the form
\begin{equation}
t(E)=\frac{\lam_1^2|\phi\rangle\langle\phi|}{E+i0-E_D-J_1(E)},
\label{te}
\end{equation}
where the function $J_1(E)$ is proportional to the matrix element of the resolvent for the external Hamiltonian
 $g_{NN}(E)\equiv [E+i0-h_{NN}]^{-1}$:
\begin{equation}
J_1(E)=\lam_1^2\langle \phi|g_{NN}(E)|\phi\rangle.
\end{equation}
Note that the imaginary part of this function can be found explicitly:
\begin{equation}
{\rm Im} J_1(E)= -\pi\lam_1^2|\langle \phi|\psi(E)\rangle|^2,
\end{equation}
where $| \psi(E)\rangle$ is the scattering function for the Hamiltonian
$h_{NN}$.

Using the expression for the transition operator (\ref {te}), it is easy to obtain the formula for the total $S$ matrix:
\begin{equation}
S(E)=e^{2i\de_0}\frac{E-E_D-\RE J_1(E)+i \IM J_1(E) }{E-E_D-\RE
J_1(E)-i \IM J_1(E)}, \label{smat}
\end{equation}
where $\de_0$ is the phase shift of the $NN$ scattering governed by the OPE poterntial.

In turn, from the formula (\ref{smat}), one finds a factor proportional to the reaction cross section:
\begin{equation}
1-|S(E)|^2=\frac{-2\IM J_1(E)\Gamma_{\rm inel}}{(E-E_0-\RE
J_1(E))^2+\quat (\Gamma_{\rm inel}-2\IM J_1(E))^2}.
\end{equation}
This gives the condition for the renormalized position of the dressed dibaryon resonance $E_R$ with respect to the bare value $E_0$\footnote{It would be more consistent to call the dibaryon in the internal channel `` semi-dressed '' since the imaginary part of its eigen-energy already takes into account all inelastic decay modes. However in our model we investigate the dressing of the dibaryon due to the connection with the $NN$ channel and from this viewpoint, the initial dibaryon is ``bare''.}:
\begin{equation}
E_R=E_0+\RE J_1(E_R), \label{res_con}
\end{equation}
and also the expression for the renormalized energy-dependent full width:
\begin{equation}
\Gamma(E)=\Gamma_{\rm inel}-2\IM J_1(E) = \Gamma_{\rm inel}+2\pi \lam_1^2|\langle
\phi|\psi(E)\rangle|^2. \label{res_con_im}
\end{equation}
After finding the position of the dressed dibaryon $E_R$ from the condition
(\ref{res_con}), the total resonance width $\Gamma_{\rm th}$ is determined from Eq.~(\ref{res_con_im}) at $E =E_R$.
The mass of the dressed dibaryon $M_{\rm th} $ is related to the energy $E_R$ by the relation $M_{\rm th} = 2\sqrt{m(E_R + m)}$, where $m$ is the nucleon mass\footnote {Here we use the method of ``minimal'' accounting for relativistic corrections (see, e.g., the work \cite{Geramb}), in which the relationship $E = k^2/m$ between the energy $E$ and the relative momentum of the pair of nucleons $k$ is conserved. However, the momentum $k$ is determined from the laboratory collision energy $T_{\rm lab}$ in accordance with the relativistic formula $k =\sqrt{mT_{\rm lab}/2}$. Then $ T_{\rm lab} = 2E$ and the full invariant energy is $\sqrt{s} = 2\sqrt{m(E + m)}$.}.

Thus, the function $ J_1 (E) $ describes shifts of the real and imaginary parts of the complex energy of the dibaryon. This shift of the resonance position occurs due to the coupling of the initial (bare) dibaryon with the external $NN$ channel. In this case, it is possible to estimate the branching ratio for the dibaryon decay into the $NN$ channel:
\begin{equation}
w = \frac{\Gamma_{\rm th} - \Gamma_{\rm inel}}{\Gamma_{\rm th}} = \frac{2\pi \lam_1^2|\langle
\phi|\psi(E_R)\rangle|^2}{\Gamma(E_R)}. \label{weight}
\end{equation}

To effectively take into account inelastic processes, as well as to describe the threshold behavior of the reaction cross section, one should introduce the energy dependence of the bare-resonance width $\Gamma_{\rm inel}$.
The main inelastic process for the isovector $NN$ channels considered in this paper is one-pion production. In turn, the three-particle channel $D\to \pi NN $  dominates in the decay width of isovector dibaryons while the two-particle channel $D\to \pi d$ makes the contribution $\lesssim 30$ \%~\cite{Strak91} and has a similar threshold behavior. The respective inelastic decay width of the dibaryon can be represented as follows:
\begin{equation}
\Gamma_D(\sqrt{s})=\left\{
\begin{array}{lr}
0,& \sqrt{s}\leq E_{\rm thr};\\\displaystyle
\Gamma_0\frac{F(\sqrt{s})}{F(M_0)},&\sqrt{s}>E_{\rm thr}\\
\end{array}\label{gamd}
\right.,
\end{equation}
where $\sqrt{s}$ is the total invariant energy of the decaying
resonance, $M_0$ the bare dibaryon mass, $E_{\rm
thr}=2m+m_\pi$ the threshold energy, and $\Gamma_0$ defines
the decay width at the resonance pole.

The function $F(\sqrt{s})$, which takes into account the dibaryon decay into the channel $\pi NN $ where the emitted pion has the orbital angular momentum $l_{\pi}$ and the $NN$ pair has the orbital angular momentum $L_{NN}$, can be parameterized using the empirical formula:
\begin{equation}
F(\sqrt{s})=\frac{1}{s}\int_{2m}^{\sqrt{s}-m_{\pi}}dM_{NN}
\frac{q^{2l_\pi+1}k^{2L_{NN}+1}}{(q^2+\Lam^2)^{l_\pi+1}(k^2+\Lam^2)^{L_{NN}+1}},
\label{fpinn}
\end{equation}
where $\displaystyle
q={\sqrt{(s-m^2_\pi-M^2_{NN})^2-4m_\pi^2M_{NN}^2}}\Big/{2\sqrt{s}}$
is the pion momentum in the total center-of-mass frame,
$\displaystyle k=\half\sqrt{M_{NN}^2-4m^2}$ the momentum of
the nucleon in the center-of-mass frame of the final $NN$
subsystem with the invariant mass $M_{NN}$, and $\Lam$ the
high-momentum cutoff parameter which prevents an unphysical rise
of the width $\Gamma_{\rm inel}$ at high energies. The orbital
momenta $l_{\pi}$ and $L_{NN}$ may take different values however
their sum is restricted by the total angular momentum and parity
conservation.
The specific values of the parameters $l_{\pi}$, $L_{NN}$ and $\Lam$
were adjusted to get an optimal description of the scattering phase shifts in given partial $NN$ channel (see Sec. 4 and 5).
It is important to note that these parameters primarily affect the threshold behavior of the inelastic phase shifts and are used for the ``fine tuning'' of the model, while the main results presented below in the paper are sensitive mainly to the mass and width of the bare dibaryon.

Thus, we have formulated here a simple model for coupling between the
external $NN$  channel (in our case driven by OPE interaction) and the internal (dibaryon) channel and have shown that the complex energy of the initial bare dibaryon is renormalized due to coupling with the $NN$ channel, resulting in the mass and width of the fully dressed dibaryon. Graphically, this interaction mechanism can be represented as a sequence of diagrams shown in Fig.~3. Such a series of diagrams actually corresponds to the Dyson equation for a dressed particle in the quantum field theory.

\bigskip
\begin{center}
4. ISOVECTOR $NN$ CHANNELS WITHOUT THE REPULSIVE CORE
\end{center}

In this section, we consider as an example the isovector
partial waves ($^3P_2$, $^1D_2$, and $^3F_3$) of $NN$
scattering, where the empirical phase shifts do not manifest
explicitly the behaviour inherent to a repulsive core at short distances, at least up to relatively high collision energies $T_{\rm lab} \simeq
800$~MeV.\footnote{The fact that at higher energies the real phase shifts in these channels become negative is a consequence of the appearance of
a large imaginary part of the phase shifts, which can be interpreted
as a strong repulsion.}
These channels were also chosen for the reason that the existence of dibaryon resonances was quite reliably established in them. The parameters of these resonances are known from experiments and can be used to test our model.\footnote{
Isoscalar channels (in particular, the channel $^3D_3$, in which the existence of the dibaryon resonance $d^*(2380)$ is well established~\cite{Bashkanov09, Adlarson11, Adlarson14}) will be considered in our next work.}

In fact, the most important point in the results presented below is {\em close agreement of the masses and widths of the dressed dibaryons obtained by fitting the $NN$-scattering phase shifts in our model with the parameters of the experimentally found dibaryons in the corresponding partial channels}
(see
\cite{Meshcher,Komarov,Auer1,Auer2,Hoshiz,Nagata,Arndt,Strak,Strak91}).

To represent the partial-wave amplitudes, we use below the parametrization for the $K$ matrix by Arndt et al. \cite{SAID2007,Geramb}, which in the case of single-channel scattering has a simple form:
\begin{equation}
K=\tan\delta +i \tan^2\rho,
\end{equation}
where $\delta$ is the real phase shift and $\rho$ is a parameter
related to inelasticity. For the sake of simplicity, below we will
refer to the parameter $\rho$ as the imaginary phase shift.

\bigskip
\begin{center}
\em 4.1. Channel $^3P_2$
\end{center}

The empirical $NN$ phase shifts in the channel $^3P_2$ as was 
found in the partial-wave analysis of the George Washington University group (SAID)
 do not display any sign of the repulsive core and
remain to be positive at least up to energies of $T_{\rm lab}
\simeq 1000$ MeV. This clearly indicates that in this channel the traditional repulsive core does not play a decisive role and the main mechanism of $NN$ interaction corresponds to a rather strong attraction.

We tried to reproduce this attraction by a single dibaryon pole in the effective Hamiltonian of $NN$ interaction (\ref{Heff}), by varying the complex energy of the dibaryon $E_D$ and the real coupling constant $\lam_1$ of the external and internal channels.

The real partial phase shifts $\delta$ and the parameters $\rho$ for the channel $^3P_2$ are displayed  in Fig.~5 in comparison with the SAID data~\cite{SAID2007}.\footnote{
The triplet channel $^3P_2$ is coupled by tensor interaction with the channel $^3F_2$. However, in the energy range under consideration (up to 700 MeV), the real and imaginary
$^3F_2$ phase shifts, as well as the mixing angle $\epsilon_2$ are small, therefore this coupling is not taken into account here.}

\begin{figure}[h!]
\centering\epsfig{file=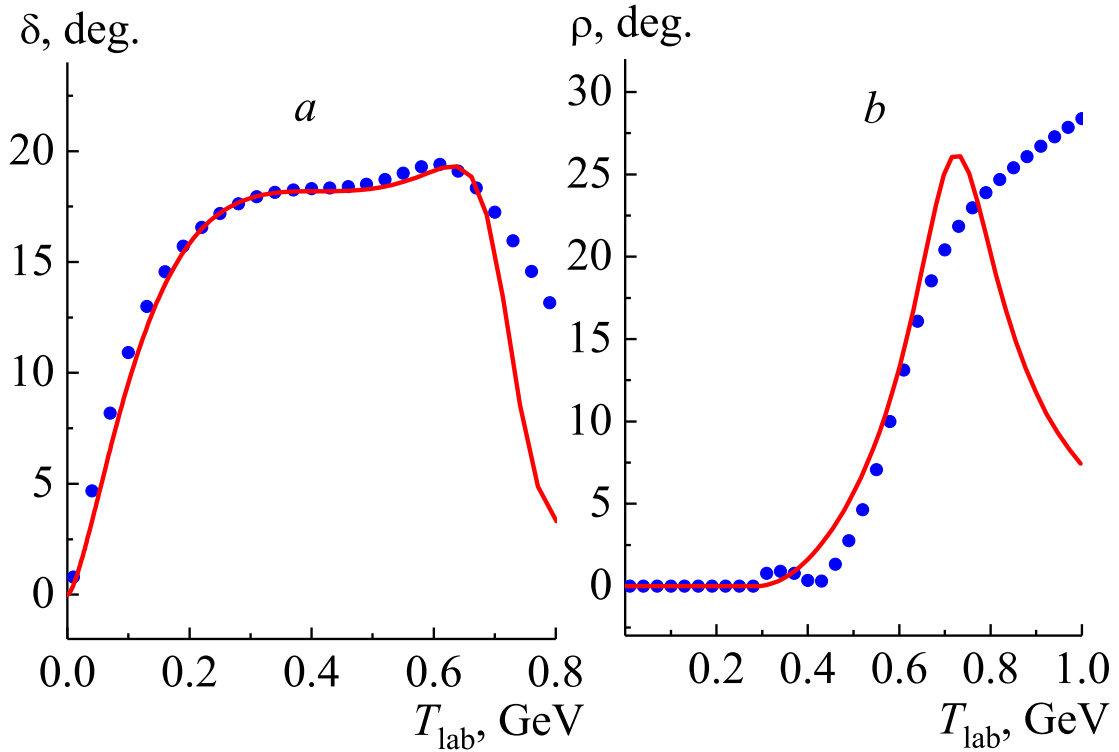,width=0.9\columnwidth}
\caption{Real $\delta$ ($a$) and imaginary $\rho$ ($b$) partial phase shifts in the channel $^3P_2$ found in the dibaryon model
(solid curves) in comparison with the PWA data of the SAID group (points).}
\label{fig5}
\end{figure}

In these calculations, we used the following parameters of the effective potential (\ref{Heff}): $\lam_1 = 0.065$~GeV and $r_0 = 0.71$~fm.
The total width of the dibaryon was parameterized in the form (\ref{gamd}), (\ref{fpinn}) with the parameters:
$\Gamma_0=0.096$ GeV, $l_\pi=2$, $L_{NN}=0$ к $\Lam=0.3$ GeV/$c$.

It can be seen from the figure that, although our model leads to somewhat overestimated imaginary phase shifts $\rho$, it still allows to describe quite well the real and imaginary phase shifts up to energies
 $T_{\rm lab} \sim 0.6 $~GeV. It is also important to emphasize here that the description of both real and imaginary phase shifts in this channel was obtained with the same parameters of the bare dibaryon.

Now using explicitly the condition (\ref{res_con}) and Eq. (\ref{res_con_im}), we found the following parameter values for the dressed dibaryon in the channel $^3P_2$:
\begin{equation}
M_{\rm th}(^3P_2)=2.23\quad{\mbox{GeV}}, \quad \Gamma_{\rm
th}(^3P_2)= 0.15\quad{\mbox{GeV}}.
\end{equation}

These values should be compared with the
respective experimental values found recently by the ANKE-COSY
Collaboration \cite{Komarov}\footnote{%
The numbers in
parentheses denote the uncertainty in the last figure}:
\begin{equation}
M_{\rm exp}(^3P_2)=2.197(8)\quad{\mbox{GeV}}, \quad \Gamma_{\rm
exp}(^3P_2)=0.130(21)\quad{\mbox{GeV}}.
\end{equation}

As one can see, the mass and width of the dressed dibaryon in the $^3P_2$ channel turn out to be very close to the mass and width of the experimentally found $^3P_2$ dibaryon (taking into account the experimental errors).

Since we made no other assumptions about the nature of interaction in this channel, except for the obvious presence of a one-pion exchange, whose contribution is rather small, the results obtained indicate directly the dominant dibaryon mechanism of the basic $NN$ interaction in this channel.

It is important to note here that a similar conclusion about the dominance of the dibaryon mechanism in inelastic $NN$ interaction in the $^3P_2 $ channel was made by us when analyzing the single-pion production reaction
$pp\to d\pi^+ $ at energies from the threshold to about 800 MeV~\cite{PRD2016}.
In the framework of a phenomenological model that takes into account both traditional meson-exchange mechanisms (with moderate values of the cutoff parameters in meson-baryon vertices consistent with the $\pi N $ scattering data) and dibaryon contributions, it was shown that isovector dibaryon resonances in leading partial waves for this reaction ($^1D_2 $, $^3F_3 $, and $^3P_2 $) play a very important role in describing the main characteristics of the reaction $pp\to d\pi^+$, including the behavior of complicated polarization observables.

Moreover, the interaction in the channel $^3P_2$
is almost completely determined by the dibaryon contribution, while the traditional mechanism of the $\Delta$-isobar excitation plays a very moderate role in this channel.\footnote{For the final states other than $d\pi$, the contribution of the $\Delta$-isobar excitation may be larger, but this does not change the qualitative conclusions about the role of the dibaryon mechanism.}
It is important to emphasize that the parameters of the $^3P_2 $ dibaryon that were found in Ref.~\cite{PRD2016} turned out to be also very close to the experimental ones~\cite{Komarov}.
Thus, the results of the present work not only confirm the conclusion that the $^3P_2 $ dibaryon dominates the inelastic $NN$ interaction up to energies of $ T_{\rm lab}\simeq 600 $ MeV, but, even more importantly, they allow us to make the conclusion about the decisive role of this resonance also in elastic $NN$ scattering in the same energy region.

\bigskip
\begin{center}
\em 4.2. Channel $^1D_2$
\end{center}

A completely similar consideration of $NN$ scattering in the channel $^1D_2$ leads to the results shown in Figs. 6$a$ and 6$b$ for the real and imaginary phase shifts, respectively. In these calculations, the potential parameters
$\lam_1 = 0.048$~GeV and $r_0 = 0.82$~fm were used. The parameters for the width of the bare dibaryon were chosen as follows: $\Gamma_0 = 0.08$~GeV, $l_\pi = 0$, $L_{NN} = 1$ and $\Lam = 0.2$~GeV/$c$.

\begin{figure}[h!]
\centering\epsfig{file=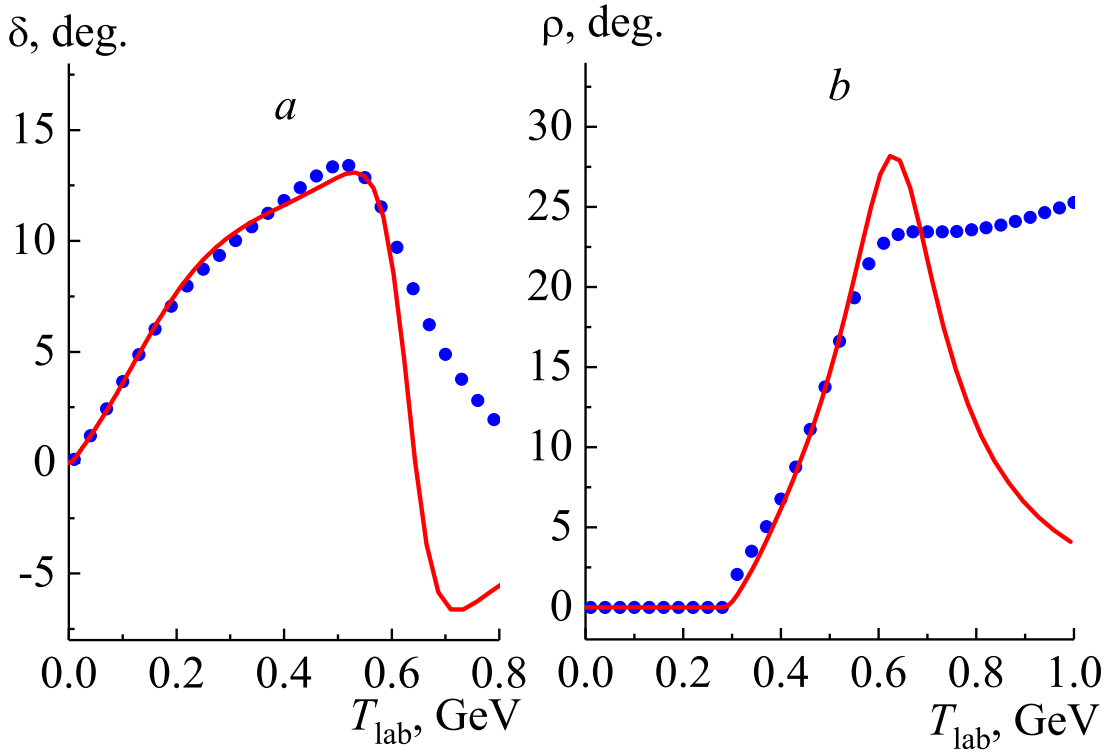,width=0.9\columnwidth}
\caption{ The same as in Fig. 5, but for the channel $^1D_2$.}
\label{fig6}
\end{figure}

In the channel $^1D_2$, the imaginary phase shifts turn out to be again a little
overestimated near the position of the dibaryon pole. However their
behavior from the inelastic threshold to energies of about 550 MeV, as well as
behavior of the real phase shifts at energies from zero to 600
MeV is reproduced almost quantitatively.

From the conditions (\ref{res_con}) and (\ref{res_con_im}), we obtained
the following parameters of the dressed dibaryon in the channel $^1D_2$:
\begin{equation}
M_{\rm th}(^1D_2) = 2.18 \quad {\mbox{GeV}}, \quad \Gamma_{\rm
th}(^1D_2) = 0.11 \quad {\mbox{GeV}},
\end{equation}
which also turned out to be rather close to the experimental values
found in a number of works~\cite{Meshcher,Auer2,Hoshiz,Nagata,Arndt,Strak} (see also summary tables of dibaryon parameters in the review \cite{Strak91}).

It is worth noting that, as has long been known,
there is a very strong coupling between $NN$ and $N\Delta$ channels in the $^1D_2$ $NN$ partial wave, since the $N\Delta$ system is produced here in the relative $S$ wave (in
the $^5S_2$ state), and this coupling determines most
inelasticity in this partial channel. Since we have effectively (through the width
of the initial ``half-dressed'' dibaryon) taken into account the resonance-type ($s$-channel) coupling with the $N\Delta$ state, the behaviour of the
inelastic phase shifts is reproduced only up to collision energies
corresponding to the mass of the $^1D_2$ dibaryon.
At higher energies, the connection between the $NN$ and $N\Delta$ states is mainly due to
the $t$-channel one-pion exchange, so that inelastic scattering above $T_{\rm lab}
\simeq 600$~MeV is no longer described by our
simple model. However, an almost quantitative description of both
real (elastic) and imaginary (inelastic) phase
shifts (with the same parameters of the initial dibaryon)
up to energies of $T_{\rm lab} \simeq 600$ MeV clearly indicates
the dominance of the dibaryon interaction mechanism in this
partial-wave channel, at least in this energy region.

The important role of the $^1D_2$ dibaryon was previously established by the present authors in
reactions of one- and two-pion production in
$NN$ collisions~\cite{NPA2016,PRD2016}. In particular, it was
shown that, although the $t$-channel excitation of the $N\Delta$ system
gives a significant contribution to this partial channel, the $pp \to d \pi^+$
reaction cross section in the vicinity of the
resonance peak ($T_{\rm lab} \sim 600$ MeV) still cannot be quantitatively reproduced without taking into account the intermediate dibaryon excitation. We emphasize here
that the results of this work demonstrate a decisive influence of the
dibaryon not only near the mass shell, but also far beyond it, both for elastic and inelastic $NN$ interactions.

\bigskip
\begin{center}
\em 4.3. Channel $^3F_3$
\end{center}

In the isovector channel $^3F_3$ the presence of a dibaryon resonance with a
mass $M({}^3F_3)\simeq 2.26$~GeV was established experimentally
in 1977 \cite{Auer1}. A very large inelasticity was also found in
this $NN$ channel. Here again
an interesting question arises: to what extent does the
$^3F_3$ resonance affects elastic and inelastic scattering in this
channel? At first glance, it seems that a noticeable contribution of the dibaryon
can be seen mainly at energies close to the resonance position. However, generally speaking, the influence of the $s$-channel dibaryon exchange in $NN$ scattering must be traced far
from the mass shell --- and we have already seen this in other
partial waves.

\begin{figure}[h!]
\centering\epsfig{file=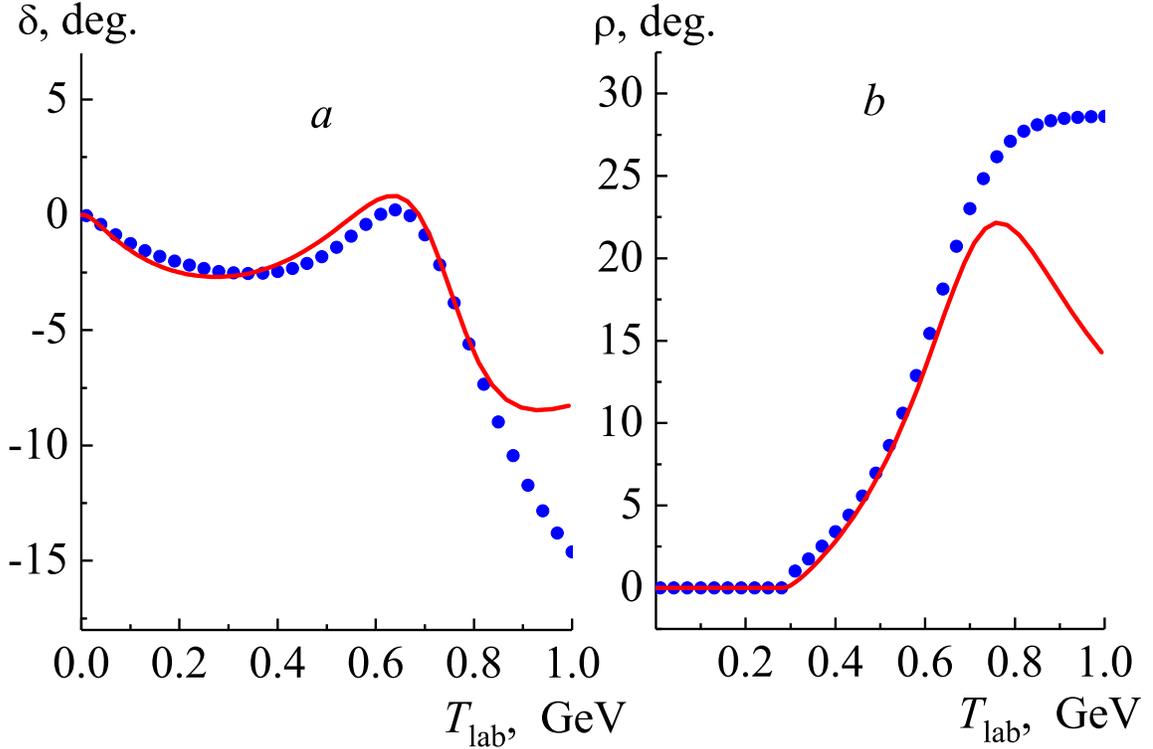,width=0.9\columnwidth}
\caption{The same as in Fig. 5, but for the channel $^3F_3$.}
\label{fig7}
\end{figure}

In Figs.~7$a$ and 7$b$ the real and imaginary
parts of phase shifts in the channel $^3F_3$ in comparison with the PWA (SAID) data are shown. Here we used again the energy-dependent resonance width (\ref{gamd}), (\ref{fpinn}) with parameters
$\Gamma_0 = 0.15$ GeV, $l_{\pi} = 0$,
$L_{NN} = 2$ and $\Lam = 0.1$ GeV/$c$. The following
potential parameters were taken: $\lam_1 = 0.065$ GeV and $r_0 = 0.5$ fm.

From Fig.~7 it is clearly seen that the model with a
single dibaryon pole (in combination with a simple
OPE interaction in the external channel\footnote{Since at
low energies the phase shift $\delta$ is very small and completely determined
by the OPE potential, it is extremely sensitive here to
the cutoff parameter $\Lam_{\pi NN}$. In calculations for the channel $^3F_3$
the value $\Lam_{\pi NN} = 0.5$ GeV was used.}) reproduces
almost quantitatively elastic phase shifts in the channel $^3F_3$ at energies from zero to
800 MeV. The behavior of the imaginary phase shift is also well reproduced
from the inelastic threshold up to energies of the order of 700
MeV. Moreover, in this channel, imaginary phase shifts near the
resonance pole are not overestimated as in other
channels considered. These results clearly indicate the dominance
of the dibaryon mechanism in elastic and inelastic $NN$ interaction in
the channel $^3F_3$ at energies up to $T_{\rm lab} \simeq
700$--$800$ MeV, and the dibaryon here manifests itself very
clearly (in particular, resonance behavior can be seen even in
the elastic phase shift).

For the $^3F_3$-dibaryon, we obtained again good agreement of the mass and
width found in our model with their experimental
values~\cite{Auer1,Auer2,Hoshiz,Nagata,Arndt,Strak,Strak91} (although
existing data have a rather large spread). Our results
for the dibaryon parameters presented in this section in
comparison with experimental data are summarized in
Table~1.
\begin{table}[h!]
\caption{ Parameters of the ``bare'' ($M_0$, $\Gamma_0$) and dressed ($M_{\rm
th}$, $\Gamma_{\rm th}$) dibaryons (in GeV) for three isovector
$NN$ channels in comparison with experimental values taken from Refs.~\cite{Komarov} ($^3P_2$) and \cite{Strak91}
($^1D_2$, $^3F_3$)}
\label{Tab1}
\begin{center}\begin{tabular}{c| c |c |c |c |c |c}\hline ${}^{2S+1}L_J$& $M_0$&
$\Gamma_0$&$M_{\rm th}$&$\Gamma_{\rm th}$&$M_{\rm exp}$&$\Gamma_{\rm exp}$\\
\hline
$^3P_2$& 2.21 &0.096 &2.23&0.15& 2.197(8)&0.130(21) \\
$^1D_2$& 2.168 &0.08 &2.18&0.11& 2.14--2.18 &0.05--0.1 \\
$^3F_3$&2.23  &0.15 &2.22&0.17&  2.20--2.26 & 0.1--0.2 \\
\hline
\end{tabular}\end{center}
\end{table}

It is also important to note that for all three
above dibaryons, the branching ratio of the decay into the $NN$ channel
defined by the formula (\ref{weight}) is 10--20\%, which
is consistent with the previous
estimates~\cite{Hoshiz,Nagata,Arndt,Strak,Strak91}.

Thus, we found that in all isovector
channels considered, generation of an intermediate (experimentally established)
dibaryon largely determines both elastic and
inelastic $NN$ scattering at energies from zero to 600--800 MeV, i.e., far from the region where the respective resonance reaches its mass shell.

\bigskip
\begin{center}
5. CHANNELS WITH REPULSIVE CORE: CHANNEL $^1S_0$
\end{center}

In this channel, as is well known, the repulsive
core at small distances clearly manifests itself, which leads to negative phase
shifts starting from the collision energies $T_{\rm lab} \simeq 250$ MeV.
Dibaryon model for $S$-wave $NN$ interaction
predicts a dominant six-quark configuration with
mixed symmetry $|s^4p^2[42]_xLST\rangle$, with two
oscillatory excitation quanta \cite{YaF2001,JPhys2001,AnnPhys2010}.
When projecting this six-quark configuration onto the $NN$ channel,
there appears a function of the relative motion of two nucleons $\chi(R)$
with an internal node. It is very important this node turns out to be
almost in the same place ($r_{NN} \sim 0.5$ fm) as the
traditional repulsive core, and, in addition, it practically does not
change its position with increasing collision energy (see
Fig.~1). So this stationary node reflects very precisely
the effects of the strong repulsive core in $NN$ scattering.

In the dibaryon model \cite{YaF2001,JPhys2001,AnnPhys2010}, appearance of such a
stationary node is provided by an additional projector in
the $NN$ potential having the form
\begin{equation}
V_{\rm rep}=\lam|\phi_0\rangle\langle \phi_0|, \label{v_rep}
\end{equation}
where $\lam \to \infty $ and $|\phi_0\rangle\langle\phi_0|$ ---
the projector onto the fully symmetric six-quark configuration
$|s^6[6]\rangle$, excluded from the model space due to its
obvious smallness compared to the dominant configuration with mixed
symmetry. Thus, in case of $S$ and  some $P$ waves ($^3P_0$,
$^3P_1$ and $^1P_1$), where the repulsive core is clearly manifested,
the effective Hamiltonian (\ref{Heff}) should be supplemented
by an orthogonalizing pseudopotential (\ref{v_rep}) with a large
positive coupling constant $\lam$.

Fortunately, this modification does not add any
free parameters to our model, so the number of adjustable parameters
remains minimal. As a result, varying the complex energy of only
one dibaryon pole, we find quite satisfactory
description of both real and imaginary parts of the $NN$-scattering phase shift
in the channel $^1S_0$ in a very broad energy
range --- from zero until 1200 MeV (see Figs.~8$a$ and 8$b$).

\begin{figure}[h!]
\centering\epsfig{file=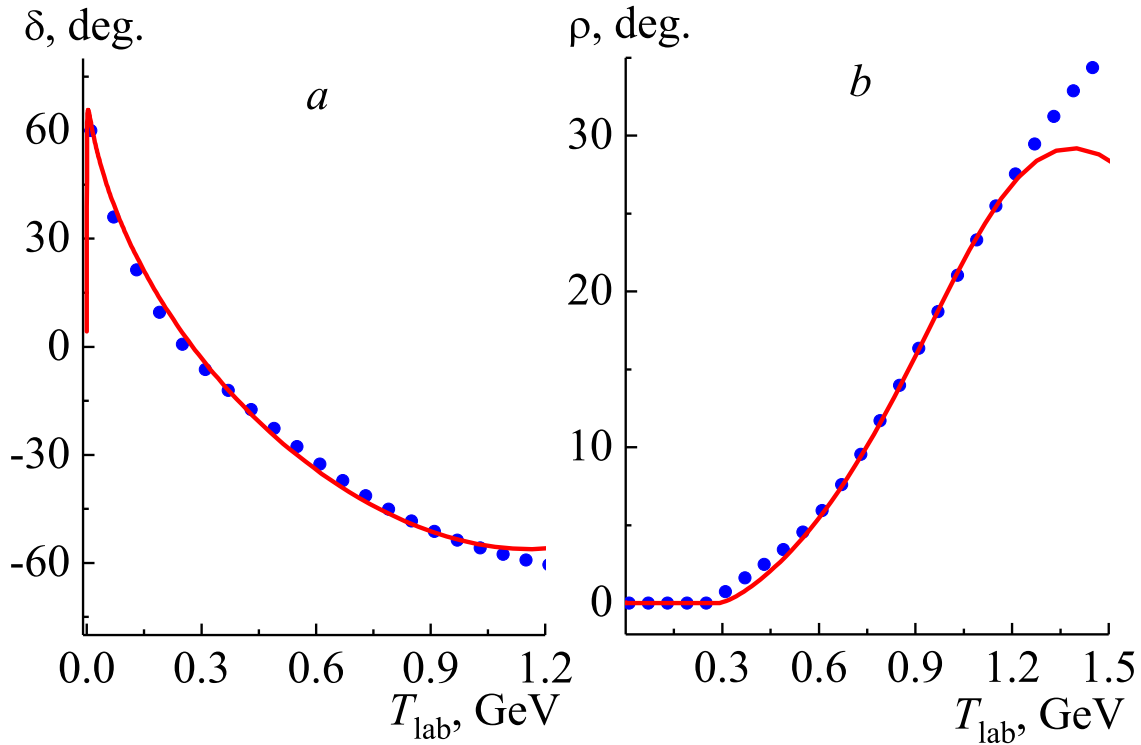,width=0.9\columnwidth}
\caption{The same as in Fig. 5, but for the channel $^1S_0$.}
\label{fig8}
\end{figure}

An amazing fact clearly follows from the results presented:
as in the cases discussed above (but in an even larger
energy interval!), description of singlet phase shifts in the channel
$^1S_0$ is basically determined by only one dibaryon pole.
However, the dressed dibaryon in this channel turns out to be very broad.
Therefore, it is more correct to determine its position not from the condition
(\ref{res_con}) on the real axis, but from the condition (\ref{smat}) for the $S$-matrix pole
in the complex energy plane:
\begin{equation}
Z - E_D - J_1(Z) = 0.
\end{equation}
In this case, for the ``bare'' dibaryon parameters\footnote{For calculations in this channel,
the potential parameters $\lam_1 = 1.184$ GeV, $r_0 = 0.51$
fm and dibaryon width parameters $l_{\pi} = 1$, $l_{NN} = 0$ and $\Lam = 0.4$ GeV/$c$ were used.}
$M_0 = 2.364$~GeV and $\Gamma_0 = 0.044$~GeV, we obtained the following
parameters of the dressed dibaryon:
\begin{equation}
M_{\rm th}(^1S_0)=2.59 \quad {\mbox{GeV}}, \quad \Gamma_{\rm
th}(^1S_0)=0.63 \quad {\mbox{GeV}}.
\end{equation}

It should be noted that this means the existence of a second
high-lying dibaryon in the channel $^1S_0$ in addition to the
near-threshold $^1S_0$-dibaryon (singlet deuteron),
predicted by Dyson and Xuong \cite{Dyson} many years ago.
It is interesting to note that the analysis of the $NN$-scattering $S$ matrix found by us in
this channel clearly indicates the presence of the singlet
deuteron pole near the $NN$ threshold (this pole is located at energy
ca. $-70$ keV on the non-physical sheet), which was not included into the
initial interaction potential, but appeared automatically due to
coupling of $NN$ and dibaryon channels.

\bigskip
\begin{center}
6. DISCUSSION: ROLE OF DIBARYONS IN NUCLEAR PHYSICS
\end{center}

In this section, we briefly discuss the important role that
dibaryon resonances play in nuclear physics as a whole. This is the main difference between six-quark dibaryons and other exotic multiquark states such as tetra-
and pentaquarks that do not play a significant role in nuclear
physics.

\bigskip
\begin{center}
\em 6.1. Incompressibility (density constancy) of nuclear matter
\end{center}

As is well known, the density of matter inside all nuclei, starting
approximately with $A = 25$, is the same on average, apart from
local fluctuations due to shell effects. In other
words, nuclear matter is practically
incompressible, despite the powerful forces of attraction between the nucleons,
which prevent the nucleus from falling to its constituents. Although
the present-day models of nuclei explain this incompressibility
``de facto'', i.e., in concrete calculations, the general idea of the reasons for such
behaviour of nuclear matter is still missing. The dibaryon concept can give
here a quite natural (at least qualitative)
explanation that allows to simultaneously interpret
a large number of experimental results, which have not yet found
an explanation within the framework of traditional models.

As we have shown in exact calculations of $^3$H and
$^3$He nuclei within the dibaryon model~\cite{sys3n}, owing to the
nodal behavior of radial wavefunctions in $S$ and
$P$ waves, the average kinetic energy of nucleons in the system is rising strongly (approximately three times in $3N$ nuclei~\cite{sys3n}) in comparison with that for traditional $NN$-force models. This rise at the constant total energy is accompanied by an increase
(in absolute value) of the average potential energy
of the nucleon attraction, i.e., nuclear matter turns out to be
much stronger ``heated'' and at the same time much stronger compressed by the huge attractive forces
induced by generation of the intermediate dibaryon
resonances. This means that the kinetic pressure of such matter
will be very strong, when turning off the powerful forces of attraction. But
since there is no real pressure on the surface of the nucleus (or
it is very small), then the following analogy arises with ordinary water.

It is well known that water is almost incompressible, i.e., has an almost constant
density over a large pressure range. The reason for this is
high internal pressure, that is, the pressure
that water at a given temperature would exert on the walls of the vessel, if one would remove the forces
of attraction between molecules. This internal pressure reaches at
room temperature a huge value --- more than 14000 atm. And in
reality it is almost completely compensated by the forces of attraction
between water molecules, so that the resulting pressure on the walls
is very small. Clearly, any external pressure much less than 14000
atm will almost not affect the internal state of water, which
means its practical incompressibility. A similar picture
arises in nuclear matter when considered within the framework of the
dibaryon concept. ``Internal kinetic pressure'' of
the nucleons in nuclear matter is so strong that any external pressure will
only slightly affect the state of such matter.

We note in passing that, according to available calculations, the equation
of state for matter inside a neutron star should be much
more stiff, i.e., the dependence of pressure on density should be much
steeper than traditional nuclear models
predict~\cite{neutron_stars,EOS}. Obviously the stiff equation of
state just means the hard compressibility of nuclear matter.

The question arises: is it possible to confirm somehow this conclusion
experimentally? As an answer, we give here very bright
results from recent experiments performed by
Karnaukhov et al. at JINR \cite{Karnaukhov}, on fragmentation
of different nuclei into fragments from $^9$Be to $^{36}$Ar by
a high-energy deuteron beam (see Figs.~9$a$--9$d$, where
data are shown for nuclear fragments from $^9$Be to $^{20}$Ne).

\begin{figure}[h!]
\centering\epsfig{file=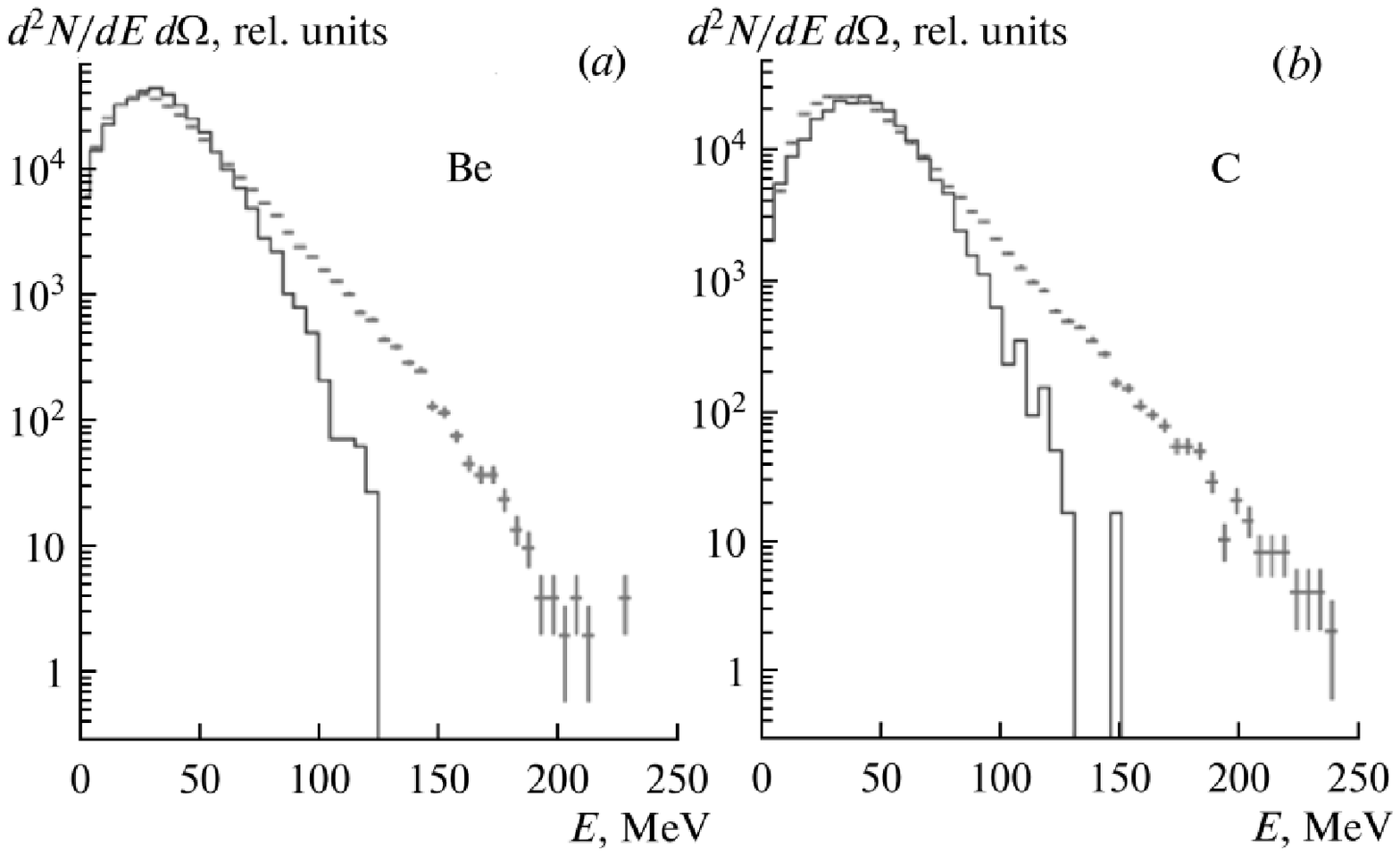,width=0.85\columnwidth}
\centering\epsfig{file=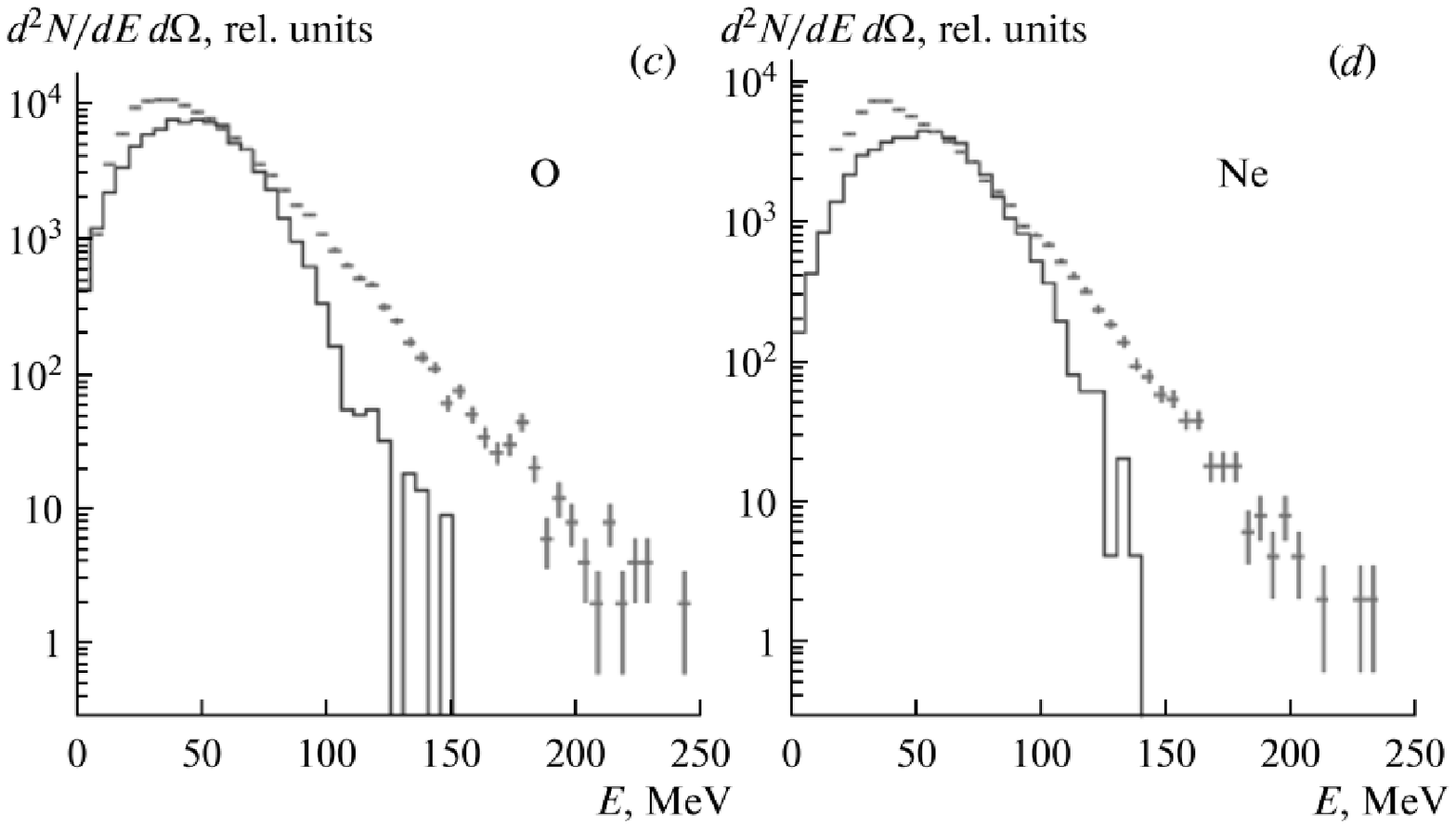,width=0.85\columnwidth}
\caption{Energy spectra of various nuclear
fragments emitted in the transverse plane (in c.m.s.) from
a target nucleus bombarded by fast deuterons. Crosses
--- data \cite{Karnaukhov} for nuclei of beryllium ($a$),
carbon ($b$), oxygen ($c$) and neon ($g$). Solid curves
--- results of theoretical calculations taking into account both
evaporative and high-momentum processes. Figure taken from \cite{Karnaukhov}.}
\label{fig9}
\end{figure}

These figures show the energy (momentum) distributions of the
nuclear fragments emitted from the target nucleus in the transverse plane upon bombarding with a beam of relativistic deuterons. In all the figures, it is clearly seen that the traditional theory
describes well the low-energy (evaporative) part of the
fragments' distributions and does not describe at all
their high-energy part. This means that under the action of a beam of fast
deuterons much more high-energy fragments are released in the target nucleus
than it follows from the traditional
theories of nuclear matter. These results can serve as a good
confirmation of strongly enhanced internal kinetic energy of the
nucleons in nuclei, which follows from the dibaryon model.

The saturation of the binding energy in nuclei is also explained very naturally on the basis of the dibaryon concept of nuclear forces.

\bigskip
\begin{center}
\em 6.2. Coulomb energies of isobar-analog states in nuclei
\end{center}


The hypothesis about the presence of dibaryon resonances in nuclei
with a probability of about 10\%  makes it possible
to explain the well-known Nolen--Schiffer anomaly for Coulomb energy shifts
of the isobar-analog states. This anomaly is that
experimental values of Coulomb energy differences for
isobar-analog states of nuclei throughout the periodic system of
elements are systematically overestimated by about 15\% in comparison with
theoretical estimates based on the known nuclear models
\cite{Nolen-Schiffer}. In other words, theory underestimates by
about 15\% Coulomb energy shifts for all
isobar-analog states in nuclei. This means that in fact,
the average distances in a pair of external protons (or neutrons) in
the isobar-analog states are slightly less than predicted
by the existing nuclear models.

Essentially the same paradox exists also for the Coulomb difference
of binding energies $\Delta E_{\rm C} = E_B(^3{\rm He}) - E_B(^3{\rm H})$ in a pair of the lightest
mirror isotopes $^3$He and $^3$H. Exact Faddeev calculations
give the value of the Coulomb difference $\Delta E_{\rm C} \simeq 630$~keV,
while its experimental value is 760 keV. For many years
this discrepancy remained unexplained. Recently
a plausible explanation of this paradox was proposed in
the works of the Bochum group~\cite{Gloeckle}, where it was shown that
the experimental value of $\Delta E_C$ can be obtained in a rigorous
theory, if one takes the $nn$ scattering length $a_{nn} =
-18.9$~fm, which is greater (in absolute value) than the well-known
``nuclear'' $pp$ scattering length $a_{pp} = -17.3$~fm. Then
the binding energy of $^3$H becomes slightly larger than that for equal
scattering lengths $a_{nn} = a_{pp}$, and $\Delta E_{\rm C}$ increases
so that it becomes equal to its experimental value.
However, recent experiments on three-body breakup $n+d \to
(nn)_0 + p$ \cite{nd-breakup} gave a different value for the $nn$ scattering length $a_{nn} = -16.5$~fm, and if to accept this value of
$a_{nn}$, then for the above Coulomb difference we get the opposite
effect: the discrepancy with the experimental value
increases even more.

A possible way out of this situation is given by the dibaryon concept of nuclear forces presented in this work. The dibaryon model predicts that the presence of dibaryons in nuclei should inevitably lead
to the appearance of a new three-body force~\cite{sys3n} due to exchange of
a light scalar $\sigma$ meson between the dibaryon and the third
nucleon. Our precise calculations of $^3$He and
$^3$H nuclei~\cite{sys3n} have shown clearly that the new theory leads to
a quantitative explanation of the Coulomb energy difference of these mirror
nuclei, thanks to an effective decrease in the mean
distances between protons in $^3$He or between neutrons in $^3$H.
Such a mechanism leads to an increase
of the intensity of the attractive pair $pp$ ($nn$) correlation in
$^3$He ($^3$H) and, accordingly, an increase of the effective binding energy
in the $pp$ ($nn$) subsystem.

\begin{figure}[h!]
\centering\epsfig{file=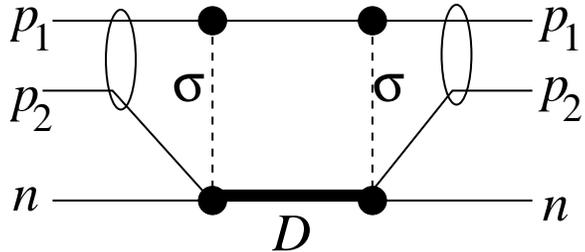,width=0.45\columnwidth}
\caption{Diagram illustrating double $\sigma$ exchange between the isoscalar ($np$) dibaryon and the second proton in $^3$He.}
\label{fig10}
\end{figure}

The diagram shown in Fig.~10 illustrates one of
the possible mechanisms for generation of such a diproton correlation in
the presence of a third particle. It is clear that such an
amplification of $pp$ correlations in nuclear medium is quite capable of explaining
fully or at least partially the Nolen--Schiffer anomaly.
We note in this respect that in the experiments of Mukha et al.~\cite{Mukha} direct diprotonic decay of a $^{94}$Ag nucleus was investigated, which is possible only when a bound diproton cluster is formed in this nucleus. This research is also closely related
to study of the contribution of dibaryon resonances to the phenomenon of
nuclear superfluidity and formation of Cooper pairs. Indeed,
strengthening of $pp$ (or $nn$) interaction in nuclear medium is
well confirmed by the appearance of superfluid pair correlations.

\bigskip
\begin{center}
7. CONCLUSION
\end{center}

The main result of this work can be formulated as follows. It has been
shown that the hypothesis of dibaryonic origin of the basic
nuclear force (at least in some $NN$-interaction channels)
allows for a reasonable and even almost
quantitative description of the behavior of real and imaginary
$NN$-scattering phase shifts over a rather broad interval
of collision energies from 0 to 600 MeV (and in the $S$ wave
--- to 1.2 GeV).\footnote{Note that the traditional so-called
realistic $NN$ potentials as well as the effective field theory
give a quantitative description of only real phase shifts in
the energy range 0--350 MeV.}
The parameters (masses and total widths) of dibaryon resonances obtained theoretically
turn out to be very close to their empirical values
found in experiments of different groups. This result allows one to
shed light on the deep connection between QCD and nuclear forces,
since the structure and dynamics of dibaryon resonances are completely
determined by the quantum chromodynamics of colored strings and quarks.

The next important result of this work is an explicit
demonstration of the fact that $NN$ interaction is highly dependent on
the specific channel and its quantum numbers (spin, isospin,
orbital angular momentum, etc.), through the dependence of the
effective $NN$-interaction potential on parameters of
the dibaryon formed in this channel and on the coupling constant with the dibaryon,
which is different for different channels. Therefore attempts to universalize
the $NN$ potential without taking into account the influence of dibaryons, by making it dependent
on general operators of the form ${\bf L}^2$, $({\bf L}{\bf S})^2$, etc.,
how this is done in realistic $NN$-potential models, lead
to a great complication of the whole interaction picture.

The results presented above clearly show that the basic
mechanism for nuclear force at small and intermediate distances, at least
in the partial-wave channels considered here, is
the formation of intermediate dibaryon resonances (dressed by meson
fields), and not a direct meson exchange between isolated
nucleons. An important confirmation of this conclusion is not only
the good description of both elastic and
inelastic $NN$ scattering demonstrated above for energies up to $T_{\rm lab} \simeq
600$--$800$~MeV (high compared to what traditional approaches give), but also satisfactory agreement of the masses and widths of dressed dibaryons obtained here with experimentally found
values.

In conclusion, the resulting picture of nuclear forces, which combines
in a unified approach both meson-nucleon and quark degrees
of freedom, makes it possible for the first time to connect fundamental
chromodynamic aspects of strong interaction (such as
restoration of chiral symmetry with increasing energy, quark confinement,
strings, etc.) and effective meson-nucleon aspects
of the problem, and thus constitutes a natural bridge
between fundamental QCD and nuclear physics. Such an approach
allows for understanding the deep interrelation between the main effects
of QCD and basic phenomena of nuclear physics, such as saturation
of nuclear forces, shell structure of nuclei, etc. The authors plan to
consider in more detail the structure of dibaryon resonances and
QCD origin of basic nuclear phenomena in the following
works.

The work was done under partial financial support of RFBR,
grants Nos.~19-02-00011 and 19-02-00014.

\end{document}